\input harvmac
\noblackbox
\newcount\figno
\figno=0
\def\fig#1#2#3{
\par\begingroup\parindent=0pt\leftskip=1cm\rightskip=1cm\parindent=0pt
\baselineskip=11pt
\global\advance\figno by 1
\midinsert
\epsfxsize=#3
\centerline{\epsfbox{#2}}	
\vskip 12pt
\centerline{{\bf Figure \the\figno:} #1}\par
\endinsert\endgroup\par}
\def\figlabel#1{\xdef#1{\the\figno}}

\def\np#1#2#3{Nucl. Phys. {\bf B#1} (#2) #3}
\def\pl#1#2#3{Phys. Lett. {\bf B#1} (#2) #3}
\def\prl#1#2#3{Phys. Rev. Lett. {\bf #1} (#2) #3}
\def\prd#1#2#3{Phys. Rev. {\bf D#1} (#2) #3}

\def\cmp#1#2#3{Comm. Math. Phys. {\bf #1} (#2) #3}


\font\cmss=cmss10
\font\cmsss=cmss10 at 7pt
\def\rlx{\relax\leavevmode}
\def\inbar{\vrule height1.5ex width.4pt depth0pt}
\def\IC{\relax\,\hbox{$\inbar\kern-.3em{\rm C}$}}
\def\IN{\relax{\rm I\kern-.18em N}}
\def\IP{\relax{\rm I\kern-.18em P}}
\def\ZZ{\rlx\leavevmode\ifmmode\mathchoice{\hbox{\cmss Z\kern-.4em Z}}
 {\hbox{\cmss Z\kern-.4em Z}}{\lower.9pt\hbox{\cmsss Z\kern-.36em Z}}
 {\lower1.2pt\hbox{\cmsss Z\kern-.36em Z}}\else{\cmss Z\kern-.4em
 Z}\fi}
\def\IZ{\relax\ifmmode\mathchoice
{\hbox{\cmss Z\kern-.4em Z}}{\hbox{\cmss Z\kern-.4em Z}}
{\lower.9pt\hbox{\cmsss Z\kern-.4em Z}}
{\lower1.2pt\hbox{\cmsss Z\kern-.4em Z}}\else{\cmss Z\kern-.4em
Z}\fi}

\def\narrowplus{\kern -.04truein + \kern -.03truein}
\def\narrowminus{- \kern -.04truein}
\def\narrowminussub{\kern -.02truein - \kern -.01truein}

\def\a{{\alpha}}
\def\g{{\gamma}}
\def\e{{\epsilon}}

\def\frac#1#2{{#1\over #2}}

\def\acom#1#2{{ \left\{ #1, #2 \right\} }}

\def\IZ{\relax\ifmmode\mathchoice
{\hbox{\cmss Z\kern-.4em Z}}{\hbox{\cmss Z\kern-.4em Z}}
{\lower.9pt\hbox{\cmsss Z\kern-.4em Z}}
{\lower1.2pt\hbox{\cmsss Z\kern-.4em Z}}\else{\cmss Z\kern-.4em
Z}\fi}
\def\IB{\relax{\rm I\kern-.18em B}}
\def\IC{{\relax\hbox{$\inbar\kern-.3em{\rm C}$}}}
\def\ID{\relax{\rm I\kern-.18em D}}
\def\IE{\relax{\rm I\kern-.18em E}}
\def\IF{\relax{\rm I\kern-.18em F}}
\def\IG{\relax\hbox{$\inbar\kern-.3em{\rm G}$}}
\def\IGa{\relax\hbox{${\rm I}\kern-.18em\Gamma$}}
\def\IH{\relax{\rm I\kern-.18em H}}
\def\II{\relax{\rm I\kern-.18em I}}
\def\IK{\relax{\rm I\kern-.18em K}}
\def\IP{\relax{\rm I\kern-.18em P}}

\font\cmss=cmss10 \font\cmsss=cmss10 at 7pt
\def\IR{\relax{\rm I\kern-.18em R}}

\def\f{\psi}

\def\dd{ {\delta}}

\def\1{{\bf 1}}
\def\3{{\bf 3}}
\def\7{{\bf 7}}
\def\2{{\bf 2}}
\def\8{{\bf 8}}

\def\W{{\cal W}}
%

%
%
\def\eqnn#1{\xdef #1{(\secsym\the\meqno)}\writedef{#1\leftbracket#1}%
\global\advance\meqno by1\wrlabeL#1}
\def\eqna#1{\xdef #1##1{\hbox{$(\secsym\the\meqno##1)$}}
\writedef{#1\numbersign1\leftbracket#1{\numbersign1}}%
\global\advance\meqno by1\wrlabeL{#1$\{\}$}}
\def\eqn#1#2{\xdef #1{(\secsym\the\meqno)}\writedef{#1\leftbracket#1}%
\global\advance\meqno by1$$#2\eqno#1\eqlabeL#1$$}



\lref\rsegal{G. Segal and A. Selby, \cmp{177}{1996}{775}. }
\lref\rdorold{N. Dorey, V. Khoze, M. Mattis, D. Tong and S. Vandoren, 
hep-th/9703228, \np{502}{1997}{59}.}
\lref\rsen{A. Sen, hep-th/9402002, Int. J. Mod. Phys. {\bf A9} (1994) 3707;
hep-th/9402032, \pl{329}{1994}{217}. }
\lref\rcallias{C. Callias, \cmp{62}{1978}{213}\semi E. Weinberg, 
\prd{20}{1979}{936}.}
\lref\rblum{J. Blum, hep-th/9401133, \pl{333}{1994}{92}.}
\lref\rsgreen{M. B. Green and S. Sethi, hep-th/9808061.}
\lref\rbateman{H. Bateman, ed. A. Erdelyi, 
{\it Higher Transcendental Functions, vol. 2}, McGraw-Hill Book Company, 1953. }
\lref\ryi{P. Yi, hep-th/9704098, \np{505}{1997}{307}.}
\lref\rsavmark{S. Sethi and M. Stern, hep-th/9705046, \cmp{194}{1998}{675}.}
\lref\rgg{M. B. Green and M. Gutperle, hep-th/9711107, JHEP 01 (1998) 05.}
\lref\rgreg{G. Moore, N. Nekrasov and S. Shatashvili, hep-th/9803265.}
\lref\rSS{S. Sethi and L. Susskind, hep-th/9702101,
    \pl{400}{1997}{265}.}
\lref\rBS{T. Banks and N. Seiberg,
    hep-th/9702187, \np{497}{1997}{41}.}
\lref\rreview{N. Seiberg, hep-th/9705117.}
\lref\rpp{J. Polchinski and P. Pouliot, hep-th/9704029, \prd{56}{1997}{6601}.}
\lref\rds{M. Dine and N. Seiberg, hep-th/9705057, \pl{409}{1997}{239}.}
\lref\rdorey{N. Dorey, V. Khoze and M. Mattis, hep-th/9704197, 
\np{502}{1997}{94}.}
\lref\rbf{T. Banks, W. Fischler, N. Seiberg and L. Susskind, hep-th/9705190, 
\pl{408}{1997}{111}.}
\lref\rlowe{D. Lowe, hep-th/9810075.}
\lref\rhoppe{G. M. Graf and J. Hoppe, hep-th/9805080.}

\lref\rK{N. Ishibashi, H. Kawai, Y. Kitazawa and A. Tsuchiya, hep-th/9612115.}
\lref\rCallias{C. Callias, Commun. Math. Phys. {\bf 62} (1978), 213.}
\lref\rPD{J. Polchinski, hep-th/9510017, \prl{\bf 75}{1995}{47}.}
\lref\rWDB{E. Witten,  hep-th/9510135, Nucl. Phys. {\bf B460} (1996) 335.}
\lref\rSSZ{S. Sethi, M. Stern, and E. Zaslow, Nucl. Phys. {\bf B457} (1995)
484.}
\lref\rGH{J. Gauntlett and J. Harvey, Nucl. Phys. {\bf B463} 287. }
\lref\rAS{A. Sen, Phys. Rev. {\bf D53} (1996) 2874; Phys. Rev. {\bf D54} (1996)
2964.}
\lref\rWI{E. Witten, Nucl. Phys. {\bf B202} (1982) 253.}
\lref\rPKT{P. K. Townsend, Phys. Lett. {\bf B350} (1995) 184.}
\lref\rWSD{E. Witten, Nucl. Phys. {\bf B443} (1995) 85.}
\lref\rASS{A. Strominger, Nucl. Phys. {\bf B451} (1995) 96.}
\lref\rBSV{M. Bershadsky, V. Sadov, and C. Vafa, Nucl. Phys. {\bf B463}
(1996) 420.}
\lref\rBSS{L. Brink, J. H. Schwarz and J. Scherk, Nucl. Phys. {\bf B121}
(1977) 77.}
\lref\rCH{M. Claudson and M. Halpern, Nucl. Phys. {\bf B250} (1985) 689.}
\lref\rSM{B. Simon, Ann. Phys. {\bf 146} (1983), 209.}
\lref\rGJ{J. Glimm and A. Jaffe, {\sl Quantum Physics, A Functional Integral
Point of View},
Springer-Verlag (New York), 1981.}
\lref\rADD{ U. H. Danielsson, G. Ferretti, B. Sundborg, Int. J. Mod. Phys. {\bf
A11} (1996) 5463\semi   D. Kabat and P. Pouliot, Phys. Rev. Lett. {\bf 77}
(1996), 1004.}
\lref\rDKPS{ M. R. Douglas, D. Kabat, P. Pouliot and S. Shenker,
hep-th/9608024,
Nucl. Phys. {\bf B485} (1997), 85.}
\lref\rhmon{S. Sethi and M. Stern, Phys. Lett. {\bf B398} (1997), 47.}
\lref\rBHN{ B. de Wit, J. Hoppe and H. Nicolai, Nucl. Phys. {\bf B305}
(1988), 545\semi
B. de Wit, M. M. Luscher, and H. Nicolai, Nucl. Phys. {\bf B320} (1989),
135\semi
B. de Wit, V. Marquard, and H. Nicolai, Comm. Math. Phys. {\bf 128} (1990),
39.}
\lref\rT{ P. Townsend, Phys. Lett. {\bf B373} (1996) 68.}
\lref\rLS{L. Susskind, hep-th/9704080.}
\lref\rFH{J. Frohlich and J. Hoppe, hep-th/9701119.}
\lref\rAg{S. Agmon, {\it Lectures on Exponential Decay of Solutions of
Second-Order Elliptic Equations}, Princeton University Press (Princeton) 1982.}
\lref\rY{P. Yi, hep-th/9704098.}
\lref\rDLhet{ D. Lowe, hep-th/9704041.}
\lref\rqm{R. Flume, Ann. Phys. {\bf 164} (1985) 189\semi
M. Baake, P. Reinecke and V. Rittenberg, J. Math. Phys. {\bf 26} (1985) 1070.}
\lref\rbb{K. Becker and M. Becker, hep-th/9705091, \np{506}{1997}{48}\semi
K. Becker, M. Becker, J. Polchinski and A. Tseytlin, hep-th/9706072,
\prd{56}{1997}{3174}.}
\lref\rss{S. Sethi and M. Stern, hep-th/9705046. }
\lref\rpw{J. Plefka and A. Waldron, hep-th/9710104, \np{512}{1998}{460}.}
\lref\rhs{M. Halpern and C. Schwartz, hep-th/9712133.}
\lref\rlimit{N. Seiberg hep-th/9710009, \prl{79}{1997}{3577}\semi
A. Sen, hep-th/9709220.}
\lref\rentin{D.-E. Diaconescu and R. Entin, hep-th/9706059,
\prd{56}{1997}{8045}.}
\lref\rgreen{M. B. Green and M. Gutperle, hep-th/9701093, \np{498}{1997}{195}.}
\lref\rpioline{B. Pioline, hep-th/9804023.}
\lref\rgl{O. Ganor and L. Motl, hep-th/9803108.}
\lref\rds{M. Dine and N. Seiberg, hep-th/9705057, \pl{409}{1997}{209}.}
\lref\rberg{E. Bergshoeff, M. Rakowski and E. Sezgin, \pl{185}{1987}{371}.}
\lref\rBHP{M. Barrio, R. Helling and G. Polhemus, hep-th/9801189.}
\lref\rper{P. Berglund and D. Minic, hep-th/9708063, \pl{415}{1997}{122}.}
\lref\rspin{P. Kraus, hep-th/9709199, \pl{419}{1998}{73}\semi
J. Harvey, hep-th/9706039\semi
J. Morales, C. Scrucca and M. Serone, hep-th/9709063, \pl{417}{1998}{233}.}
\lref\rdinetwo{M. Dine, R. Echols and J. Gray, hep-th/9810021.}
\lref\rdineone{M. Dine, R. Echols and J. Gray, hep-th/9805007, 
\pl{444}{1998}{103}.}
\lref\rber{D. Berenstein and R. Corrado, hep-th/9702108, \pl{406}{1997}{37}.}
\lref\rnonpert{A. Sen, hep-th/9402032, \pl{329}{1994}{217}; hep-th/9402002,  
Int. J. Mod. Phys. {\bf
A9} (1994) 3707\semi
N. Seiberg and E. Witten, hep-th/9408099, \np{431}{1995}{484}.
}
\lref\rpss{S. Paban, S. Sethi and M. Stern, hep-th/9805018, 
\np{534}{1998}{137}.}
\lref\rpsst{S. Paban, S. Sethi and M. Stern, hep-th/9806028, J.H.E.P.  
(1998) 9806: 012. }
\lref\rthree{S. Paban, S. Sethi and M. Stern, hep-th/9808119.}
\lref\rgreen{M. B. Green and S. Sethi, hep-th/9808061.}
\lref\rperiwal{V. Periwal and R. von Unge, hep-th/9801121.}
\lref\rfer{M. Fabbrichesi, G. Ferreti and R. Iengo, hep-th/9806018.}
\lref\ronelowe{D. Lowe, hep-th/9810075, J.H.E.P. (1998) 9311: 09.}
\lref\rtwolowe{D. Lowe and R. von Unge, hep-th/9811017, 
J.H.E.P. (1998) 9811: 014.}
\lref\rrocek{F. Gonzalez-Rey, B. Kulik, I. Y. Park and M. Rocek, 
hep-th/9810152.}
\lref\rbuck{E. I. Buchbinder, I. L. Buchbinder and S. M. Kuzenko,
hep-th/9810239.}
\lref\rBFSS{T. Banks, W. Fischler, S. H. Shenker, and L. Susskind,
    hep-th/9610043, \prd{55}{1997}{5112}.}
\lref\rhyone{S. Hyun, Y. Kiem and H. Shin, hep-th/9808183, 
\prd{59}{1999}{21901};
 hep-th/9901105;  hep-th/9901152.}
\lref\rhytwo{S. Hyun, Y. Kiem and H. Shin, hep-th/9903022.}
\lref\rokawa{Y. Okawa and T. Yoneya, hep-th/9806108, \np{538}{1999}{67}; 
hep-th/9808188, \np{541}{1999}{163}. }
\lref\rtaylor{W. Taylor and M. Raamsdonk, hep-th/9806066, 
\pl{438}{1998}{248}\semi
R. Echols and J. Grey, hep-th/9806109\semi J. McCarthy, L. Susskind and A. 
Wilkins,
hep-th/9806136, \pl{437}{1998}{62}\semi M. Fabbrichesi, G. Ferreti and R. Iengo,
hep-th/9806166, J.H.E.P. (1998) 9806: 002.}
\lref\rarvind{M. Dine and A. Rajaraman, hep-th/9710174, \pl{425}{1998}{77}.}
\lref\rnewokawa{Y. Okawa, hep-th/9903025.}
\lref\rpio{B. Pioline, hep-th/9804023, \pl{431}{1998}{73}. }

\Title{\vbox{\hbox{hep-th/9903049}
\hbox{DUK-CGTP-99-01, IASSNS--HEP--99/17}}}
{\vbox{\centerline{Supersymmetry and the Yang-Mills Effective}
\vskip8pt\centerline{Action at Finite N}}}
\centerline{Savdeep
Sethi$^\ast$\footnote{$^1$} {sethi@sns.ias.edu} and Mark
Stern$^\spadesuit$\footnote{$^2$} {stern@math.duke.edu} }
\medskip
\medskip\centerline{$\ast$ \it School of Natural Sciences, Institute for
Advanced Study, Princeton, NJ 08540, USA}
\medskip\centerline{$\spadesuit$ \it Department of Mathematics, Duke University,  
Durham, NC 27706, USA}

\vskip 0.5in

We study the effective action of quantum mechanical $SU(N)$ Yang-Mills theories 
with sixteen supersymmetries and $N>2$. We show that supersymmetry requires that
the eight fermion terms in the supersymmetric completion of the $v^4$ terms 
be one-loop exact. 
We also show that the twelve fermion terms in the supersymmetric completion of 
the $v^6$ terms are two-loop exact for $N=3$. For $N>3$, this no longer
seems to be true; we are able to find non-renormalization theorems for only 
certain twelve fermion structures. We call these structures 
`generalized F-terms.' We argue that as the rank of the gauge group is 
increased,
there can be more generalized F-terms at higher orders in
the derivative expansion. 

\vskip 0.1in
\Date{3/99}

\newsec{Introduction }

Symmetry principles give us rather remarkable control on the low-energy
physics of supersymmetric gauge and gravity theories. For example, instanton
corrections to the $F^4$ terms in three-dimensional N=8 $SU(2)$ Yang-Mills
can be determined using just supersymmetry \rthree. Likewise, all D-instanton 
corrections
to the $R^4$ terms in the type IIB string effective action are  
determined by supersymmetry \refs{\rgreen, \rpio}. The aim of this work is to 
explore the extent 
to which supersymmetry determines the structure of the effective action of
Yang-Mills theories with maximal supersymmetry and gauge group $SU(N)$\foot{It 
is worth pointing out that most of the arguments that
we will use do not depend on the Weyl group of $SU(N)$. The results should 
therefore extend to any group with rank $N-1$.}  with
$N>2$. 
In attempting to extend the technique developed in 
\refs{\rpss, \rpsst} beyond rank one, we will find some interesting new physics
and some subtleties. 

We will study the effective action at a generic point on
the Coulomb branch where the gauge group is broken to its maximal 
torus.\foot{It would be very interesting to perform a similar analysis for the 
effective action at a singularity where part of the non-abelian
gauge group remains unbroken.} Largely for notational simplicity, we will
study the quantum mechanical gauge theory that describes the low-energy
dynamics of D0-branes \refs{\rPD, \rWDB}. This theory first appeared in 
\refs{\rCH, \rqm}. A similar analysis can be performed for 
Yang-Mills theories in higher dimensions.      

In section two, we consider terms of order $v^4$. More precisely, we study all
possible eight fermion terms in the supersymmetric completion of $v^4$. We show
that all eight fermion terms must be generated at one-loop. Our results extend
 the analysis presented in \refs{\rpss,\ronelowe} for
the quantum mechanical gauge theory. To show that the remaining terms at order 
$v^4$ are one-loop exact requires either constructing the full effective 
action using a Noether procedure, or using the arguments described in 
\refs{\rhyone, \rhytwo}. We certainly expect that all terms at 
order $v^4$ in the quantum mechanics are one-loop exact as a consequence of the 
non-renormalization of the eight fermion terms. That the four derivative terms 
are only generated at one-loop in four-dimensional Yang-Mills has been argued in 
\refs{\rds, \rrocek, \rbuck, \rtwolowe}. Note that our results are in accord 
with
expectations from Matrix theory \rBFSS. 

In section three, we study the constraints imposed by supersymmetry on the 
twelve fermion terms in the supersymmetric completion of $v^6$. We show that
these terms must be two-loop exact for $SU(3)$. This agreement explains,
in large part, why matrix theory correctly reproduces the 
three body interactions of supergravity to this order in the velocity 
expansion \refs{\rokawa, \rtaylor, \rarvind}. It is important to determine
whether the twelve fermion terms completely determine the rest of the terms
at order $v^6$ \rnewokawa. Hopefully, this can be determined using the kinds of 
arguments
developed in \refs{\rhyone, \rhytwo}. We suspect that this will be the case for 
$N=2$ and
$N=3$. 

However, for higher rank gauge
groups, we have not shown that the twelve fermion terms are two-loop
exact. Rather, we are only able to show that
special twelve fermion structures are protected by non-renormalization theorems. 
Supersymmetry does impose restrictions on the remaining possible twelve fermion 
structures. These restrictions should, for example, constrain the allowed tensor 
structures. However, it is not clear that the constraints are sufficient to 
completely
determine the coupling constant dependence. 

That certain twelve fermion terms might be renormalized beyond two-loops for 
$N>3$
agrees with the perturbative computations in \refs{\rdineone, \rdinetwo}. Based 
on
the results in \rdinetwo, we would expect arbitrary renormalizations of certain
$v^6$ structures for $N>3$. In the final section, we describe the notion of 
generalized F-terms. We show that there are possible structures at order eight 
in the 
derivative expansion which, for $N>3$, must be generalized F-terms. It seems 
likely that 
the terms found in \rdinetwo\ which agreed with supergravity are in the 
supersymmetric completion of generalized F-terms. 

\newsec{Constraining Terms With Four Derivatives}

\subsec{Grading the eight fermion terms}

The Lagrangian for the supersymmetric quantum mechanics contains
bosonic fields $x^i_A$ as well as fermions $ \f_{aA}$, where $i=1,\ldots, 9$ and 
$a=1,\ldots,16$. The label $A=1,\ldots,N-1$ denotes a particular element of the 
Cartan sub-algebra of the gauge group $G$. The Lagrangian describing the 
dynamics 
at a generic flat point
has $Spin(9) \times \W$ as a symmetry group, where $\W$ is the Weyl group of 
$G$. We will take $G$ to be $SU(N)$.    

The $Spin(9)$ Clifford algebra can be represented
by real symmetric matrices $\g^i_{ab}$, where $i=1,\ldots,9$ and
$a=1,\ldots,16$. These matrices satisfy the relation,
\eqn\clifford{ \{ \g^i, \g^j \} = 2 \delta^{ij}, }
and a complete basis contains $\left\{ I, \g^i, \g^{ij}, 
\g^{ijk}, \g^{ijkl} \right\}$, where we
define:
\eqn\defs{ \eqalign{ \g^{ij} &= {1\over 2!} ( \g^i \g^j - \g^j \g^i) \cr
\g^{ijk} &= {1\over 3!}( \g^i \g^j \g^k - \g^j \g^i \g^k + \ldots) \cr
\g^{ijkl} &= {1\over 4!}( \g^i \g^j \g^k \g^l - \g^j \g^i \g^k \g^l + \ldots).
 \cr}}
The basis decomposes into symmetric $\left\{ I, \g^i, \g^{ijkl} \right\}$,
and antisymmetric matrices $\left\{ \g^{ij}, \g^{ijk} \right\}$. 

The Lagrangian $L$ can be written as a sum of terms $L= \sum L_k$ where $L_{k}$ 
contains terms of order $2k$ in a derivative expansion. The order counts the
number of derivatives plus twice the number of fermions.  
Supersymmetry requires the metric to be flat \refs{\rpss, \ronelowe}. 
The supersymmetry transformations then take the form, 
\eqn\newtransforms{ \eqalign{ \dd x^i_A & = -i \e \g^i \f_A  + 
\e  N^i_{AB} \f_B\cr
\dd \f_{aA} &= ( \g^i v^i_A \e )_a + ( M_A \e )_a.}}
The terms $N^i$ and $M$ encode all higher derivative corrections to the 
supersymmetry transformations and $\e$ is a sixteen component 
Grassmann parameter. Note 
that once higher derivative terms appear in 
$L$, we must have $N^i$ and $M$ non-zero or the supersymmetry algebra no longer
closes. Terms of order $v^4$ which appear in $L_2$ induce corrections to the
lowest order supersymmetry transformations of order 2 in $N^i$ and order 3
in $M$. To determine the eight fermion terms, we will not need to know the 
detailed form of these corrections. 

Following the argument for the $SU(2)$ case \rpss, we can immediately conclude
that the nine fermion terms which result by varying a boson in the eight fermion
terms must vanish. We then obtain sixteen first order equations that must be 
satisfied by the eight fermion terms, 
\eqn\constraint{ \sum_{A, i, b}{\g^i_{ab} \, \f_{b A} 
{\partial \over \partial x^{i}_A}\left( f^{(8)} (x)  \right)} =0, }
where we have  schematically denoted the eight fermion terms by $ f^{(8)} ( x)$.
Our task is then to unravel the extent to which \constraint\ determines the
eight fermion terms.

It is useful to grade the eight fermion terms in the 
following way: let us pick a preferred direction in the Cartan sub-algebra, 
say $A=1$, with corresponding fermions $\f_{a1}$. Any operator containing 
fermions
can then be decomposed into pieces with fixed numbers of $\f_{a1}$. We can then 
express the eight fermion terms in the form,
\eqn\eight{ f^{(8)} ( x)  = \sum_{i=0}^{8}{ f^{(8)}_{i}(x), }}
where the eight fermion term $f^{(8)}_i(x)$ contains $i$ of our preferred 
fermions
$ \f_{a1}$. Our constraint \constraint\ implies, 
\eqn\isolateone{ \left( \sum \g^i_{ab} \, \f_{b A} 
{\partial \over \partial x^{i}_A} f^{(8)}_8  \right)
+\left( \sum \g^i_{ab} \, \f_{b 1} 
{\partial \over \partial x^{i}_1} f^{(8)}_7  \right) =0. }
After multiplying \isolateone\ on the left by $ \f_{a1}$ and summing on $a$,
we can conclude that 
\eqn\isolatetwo{ \left( \f_{1} \g^i \f_A \right)
{\partial \over \partial x^{i}_A} f^{(8)}_8 =0,}
because the second term in \isolateone\ vanishes. Since 
$f^{(8)}_8$ only contains
$ \f_1$ fermions, we obtain sixteen equations:
\eqn\isolatethree{ \sum_b \g^i_{ab} \f_{b1} {\partial \over 
\partial x^{i}_A} f^{(8)}_8 =0,}
for every $A $. Note that the case $A=1$ follows directly from \constraint. 

\subsec{The homogeneity of the eight fermion terms}

We now want to show that $ f^{(8)}_8$ is one-loop exact. The coupling constant, 
$g^2$, has mass dimension 3 in these quantum mechanical gauge theories. The 
fermions are dimension $3/2$ while the scalars are dimension 1. If $ f^{(8)}_8$
is one-loop exact, it must then be a homogeneous function of the scalars $x^i_A$
of degree $-11$. For example in the rank 1 case, some of the eight fermion
terms had the form $ g^2 \f^8 / r^{11}$. 

To show that this is again the case, let us apply $ \g^j_{ca} x^j_A ({\partial 
\over
\partial \f_{c1}}) $ 
to \isolatethree\ and sum on $a$:
\eqn\scaling{ \left( 8 \sum_i x^i_A {\partial \over \partial x^i_A} - 
x^j_A {\partial \over \partial x^i_A} \g^{ij}_{cb} {\partial \over \partial 
\f_{c1}} \f_{b1} \right) f^{(8)}_8 = 0.}
Note that we have not yet summed on the $A$ index in \scaling. The second 
term in \scaling\ contains operators that generate $Spin(9)$ rotations on the
bosons and fermions. Since $f^{(8)}$ is a term in the Lagrangian, it is 
$Spin(9)$ invariant. For $f^{(8)}_8$,  this reduces to the assertion 
that, 
\eqn\inv{ \sum_A{ \left( x^j_A {\partial \over \partial x^i_A} - 
x^i_A {\partial \over \partial x^j_A} \right) f^{(8)}_8} 
= {1\over 2} \g^{ij}_{cb}{\partial \over \partial\f_{c1}} \f_{b1} 
f^{(8)}_8.}
We can use \inv\ to rewrite the second term in \scaling\ after summing on $A$,
\eqn\cas{ \left( 8 r {\partial \over \partial r} - 
 {1\over 2}\sum_{i<j}(\g^{ij}_{cb} {\partial \over \partial 
\f_{c1}} \f_{b1})^2 \right) f^{(8)}_8 = 0,} 
where $r^2 = \sum_{i,A}(x^i_A)^2.$

The last term in \cas\ can be written as  $2\rho_1(C)$ where $C$ denotes 
the Casimir operator of $Spin(9)$ and $\rho_1$ denotes the 
representation of $Spin(9)$ obtained from the product of eight $\f_{c1}$
fermions. Since $f^{(8)}_8$ is invariant under \cas, this representation
must be a polynomial representation of $Spin(9)$. Otherwise, we could
not contract our eight fermions with scalars $x^i_A$ to get an invariant
term. To determine the homogeneity of the $ f^{(8)}_8$ term, we therefore need
to evaluate the possible values of the Casimir appearing in \cas.

\subsec{Evaluating the Casimir}

 Let us introduce some
notation for the weights of $Spin(9)$. We will choose a Cartan sub-algebra,
a Weyl chamber and an orthonormal basis of weight vectors
 $<w_1,w_2,w_3,w_4>$ for the vector representation of $Spin(9)$. 
The roots are constructed in terms of these weights and the positive roots are
$ w_i \pm w_j$ with $i<j$ and $w_i$.
With this normalization, the sum of the positive roots, $2 \delta,$ 
is given by: 
$$2\delta = 7 w_1 + 5 w_2 + 3 w_3 + w_4.$$
The ${\bf 16}$ spinor representation of $Spin(9)$ then has highest weight,
$$ {1\over 2}(w_1+w_2+w_3+w_4),$$
and all the weights of this representation are of the form ${1\over 2} (\pm w_1
\pm w_2 \pm w_3 \pm w_4)$. The product of eight fermions is the reducible
representation $ \Lambda^{8} \, {\bf 16}$. It is not hard to check that 
$\rho_1(C)$
takes its largest value on the irreducible sub-representation with highest 
weight
$4w_1$.

In fact, we will see below that the constraint equations 
force $f^{(8)}_8$ to take values in this representation.\foot{Note 
that this representation contains the four scalar,
two scalar and zero scalar terms that appeared in \rpss. Together they form
an irreducible representation of $Spin(9)$. } 
The value of the Casimir on an 
irreducible subspace of highest weight $\lambda$ is given by, 
$$<\lambda+2\delta,\lambda>,$$ 
and $2\rho_1(C)$ evaluated on $f^{(8)}_8$ then gives:  
$$2<(7+4)w_1+5w_2+3w_3+w_4, 4w_1> = 88.$$
Equation \scaling\ then becomes,  
\eqn\finalscal{ \left(  r {\partial \over \partial r} + 
11 \right) f^{(8)}_8 = 0,}
and the solution is homogeneous of degree $-11$ as claimed.

We obtain weaker harmonicity constraints from \isolatethree\ in the 
following way. Apply the operator,
$$ \g^q_{ac} ({\partial \over \partial \f_{c1}}) ({\partial \over 
\partial {x^q_A}}), $$ 
to \isolatethree\ and sum on $a$ to obtain, 
\eqn\weaker{ \sum_i {\partial^2 \over (\partial x^{i}_A)^2} \, f^{(8)}_8 =0,}
for every $A$. Moreover this result is not dependent on the choice 
of coordinates used here. This means that $f^{(8)}_8$ is harmonic when 
restricted 
to any $Spin(9)$ invariant $9$-dimensional subspace of our $9r$-dimensional
moduli space 
determined by a choice of element in the Cartan of $SU(N)$. 
Borrowing a term 
referring to a similar concept from the theory of 
several complex variables, we will call such functions pluri-harmonic.

We now show that $f^{(8)}_8$ must lie in the subspace with 
highest weight $4w_1$. This calculation will also be useful in 
our later analysis. We can choose coordinates for $Spin(9)$ so 
that $\g^{12}$ is dual to the weight vector $w_1$. Since $\g^{12}$
squares to $- {\rm I}$, we can decompose our fermions into eigenvectors
of the $1-2$ generator of $Spin(9)$ rotations on the fermions given in 
\inv,
$$ \f_{aA} = \f_{aA}^+ + \f_{aA}^-, $$
where $\f_{aA}^+$ and $\f_{aA}^-$ have eigenvalues $ +i/2$ and $-i/2$, 
respectively. Note that  $\f_{aA}^+$ and $\f_{aA}^-$ are complex conjugates.

Likewise, we can decompose the canonical momenta $p^i_A$ obeying the 
usual commutation relations, 
$$ [x^i_A, p^j_B] = i\delta^{ij} \delta_{AB},$$
into eigenvectors under the $1-2$ generator of $Spin(9)$ rotations on the bosons 
also given in \inv. In this case, $p^j_A$ for $j \neq 1,2$ is clearly 
annihilated
by the rotation generator. The remaining two momenta are conveniently written
as, 
\eqn\hol{ \eqalign{ p^1_A & = {1\over 2}\left(
\partial_{z_A} + \partial_{ \bar{z}_A} \right) \cr
 p^2_A & = -{i\over 2} \left( \partial_{z_A} - \partial_{ \bar{z}_A} \right), 
\cr}}
where $\partial_{z_A}$ and $\partial_{ \bar{z}_A}$ have eigenvalues $-i$ and 
$i$,
respectively.

With these observations, we can decompose the free supercharge,\foot{Throughout
this paper, when referring to the supercharge, we will mean the operator given
in $(2.15)$
that increases fermion number. We will never need the component that 
decreases fermion number. } 
\eqn\charge{ Q_a = \g^i_{ab} \f_{bA} \, p^i_A, }
into a sum of two operators $Q_{a}^+ + Q_{a}^-, $ where 
$Q_{a}^+$ raises the $w_1$ component of the weight by $1/2$ and 
$Q_{a}^-$  lowers the $w_1$ component of the weight by $1/2$. Note that
$Q_a^-$ is the complex conjugate of $Q_{a}^+$.   
We may further decompose  
$Q_{a}^-$ (and correspondingly $Q_{a}^+$)  into a sum 
of two operators: one which raises the fermionic $w_1$ component of the weight 
by 1/2 and therefore lowers the bosonic component by 1, and one 
which leaves the bosonic component unchanged and lowers the fermionic component 
by 1/2. With a choice of complex coordinates, the first operator 
is simply  $dz_{\a A} \partial_{z_A}$ where $\a $ runs from 1 to 8 and 
$dz_{\a A}$ is a linear combination of $\f_{aA}^+$.

The highest weight component is automatically annihilated by $Q_a^+$. 
It must also be annihilated by the operator in $Q_a^-$, 
$$dz_{\a 1} \partial_{z_1},$$ 
for all $\a$.  This implies that this highest weight component is either 
anti-holomorphic in the $z_1$ variable, 
or that it is annihilated by $dz_{\a 1}$ for all $\a$. The first condition is 
not 
possible. 

To see this, note that the eight fermion term can only be singular at a point 
where
non-abelian gauge symmetry is restored. These loci are codimension nine in the
moduli space. So as we go off to infinity in almost all directions, the eight 
fermion
term must vanish. However, any anti-holomorphic function that is bounded almost 
everywhere is constant. A constant eight fermion term is unphysical.  
We must therefore satisfy the second condition. This condition means that 
the highest weight component of 
$f^{(8)}_8$ is a multiple of  $ \prod_{\a=1}^8 dz_{\a 1}$.  
This structure has weight $4w_1$ as we desired. 

As a bonus, this argument shows that if $f^{(8)}_8 = 0$ 
then $f^{(8)} = 0$, because the highest weight term with the greatest number 
of $\f_{a1}$ factors must contain all $8$ $dz_{\a 1}$ factors. Since $f^{(8)}_8$
determines all the remaining eight fermion terms, the remaining terms must all 
be 
one-loop exact. Therefore,
the eight fermion terms are one-loop exact.

\newsec{Constraining Terms With Six Derivatives}
\subsec{The homogeneity of a special twelve fermion term}

We can now consider the twelve fermion terms in the supersymmetric completion
of $v^6$. As before, we can grade the twelve fermion terms according to the
number $i$ of $\f_{a 1}$ factors: 
\eqn\twgrad{ f^{(12)} = \sum_{i=0}^{12} f^{(12)}_i.}
Supersymmetry now requires that the thirteen fermion term obtained by varying
$f^{(12)}$ satisfy the following equations for each $a$ \rpsst, 
\eqn\secconstraint{ \sum_{A, i, b}{\g^i_{ab} \, \f_{b A} 
{\partial \over \partial x^{i}_A}\left( f^{(12)} (x)  \right)} = \delta_a L_2.}
All that we need to know about the source terms $\delta_a L_2$ is that they 
are generated by
varying terms of order $v^4$ contained in $L_2$ using corrections to the 
supersymmetry transformations,
encoded in $N$ and $M$ of \newtransforms, generated by these terms of order 
$v^4$. 
The source terms in \secconstraint\ are therefore two-loop exact, and the 
corresponding
 solution to \secconstraint\ will be the sum of a two-loop exact term and a 
solution to the associated homogeneous equation 
\eqn\secconstrainto{ \sum_{A, i, b}{\g^i_{ab} \, \f_{b A} 
{\partial \over \partial x^{i}_A}\left( f^{(12)} (x)  \right)} = 0.} 
We are then left to analyze the solutions of \secconstrainto. 
In the rank one case, there was no  solution to  \secconstrainto.\foot{There 
was a homogeneous solution to the weaker harmonicity equation. This solution  
required a negative power of the 
coupling constant and so was unphysical \rpsst.} Is this again the case for a 
higher rank gauge group? 

We begin as in the $f^{(8)}$ case by considering homogeneous $f^{(12)}_{12}$, 
 which we henceforth assume is a solution to \secconstrainto. The same argument
as before gives, 
\eqn\newisolatethree{ \sum_b \g^i_{ab} \f_{b1} {\partial \over 
\partial x^{i}_A} f^{(12)}_{12} =0,}
for every $a$ and $A$.
Equation \newisolatethree\ again implies pluri-harmonicity, and retracing the 
argument
for the eight fermion case, we again obtain a radial equation:

\eqn\radtwelve{ \left( 4 r {\partial \over \partial r} + 2\rho_1(C) \right) 
f^{(12)}_{12} = 0.}
The only difference with the previous case is that the coefficient of 
$r \partial_r$ for an $f^{(k)}_k$ term is 
given by $16 - k$.  
We are left again with evaluating the Casimir term. The representation again
must contain polynomial representations since we construct an invariant term
by contracting our fermion structure with scalars $x^i_A$. The key question
is determining an upper bound on the highest weight of the representations 
that can appear in the product
of 12 fermions. It is easy to check that the highest weight that can appear 
in the exterior product of 12 
fermions transforming in the $ {\bf 16}$ of $Spin(9)$ is $2w_1$.
So the largest value of the Casimir term is, 
$$2\rho_1(C) \leq 2 (14+4) = 36.$$
This implies that $f^{(12)}_{12}$ has homogeneity $-9$ or larger. However, this
scaling behavior corresponds to a negative power of the coupling constant and is 
therefore again ruled out. So at least  the homogeneous solution for 
$f^{(12)}_{12}$ must vanish for any $N$. 

\subsec{Determining the remaining twelve fermion terms}

Unlike the eight fermion case, it no longer follows readily 
that if $f^{(12)}_{12}=0$, the rest of the $f^{(12)}_{i}$ terms must vanish. 
In fact, we have found that there are twelve fermion structures that can 
potentially
be renormalized beyond two-loops for $N>3$. For the rest of this section, we
will restrict to the $N=3$ case where we will show that the twelve fermion 
terms cannot be renormalized beyond two-loops.   

 From the previous discussion, we know that the homogeneous solution for 
$f^{(12)}_{12}$ must vanish for any
$N$. Let us take  $f^{(12)}_{i_{max}}$ as the $f^{12}_{i}$ term with largest
$i$ which
is non-zero. What follows immediately from \secconstraint\ with the source term
zero is that 
$f^{(12)}_{i_{max}}$ is harmonic in the $x_1$ direction. From \secconstraint, 
we can still deduce that
\eqn\isofour{ \left( \f_{1} \g^i \f_A \right)
{\partial \over \partial x^{i}_A} f^{(12)}_{i_{max}} =0,}
but this equation no longer implies a relation analogous to \newisolatethree. 

We will therefore have to use a different strategy to constrain the remaining
possible twelve fermion terms. Since $f^{(12)}_{i_{max}}$ is killed by 
$\Delta_1$, we can reduce $i_{max}$ by applying $\Delta_1$ to $ f^{(12)}$. 
Moreover, this `new' twelve fermion term is still killed by each supercharge
$Q_a$. Now applying a Laplacian to the twelve fermion term only makes it decay
more quickly. So this modification decreases the homogeneity. This procedure
should therefore give us a lower bound on the homogeneity of $ f^{(12)}$. 
We can keep 
repeating this procedure until all terms in the resulting twelve fermion term
are killed by  $\Delta_1$.     

Having applied this procedure in the $A=1$ direction, we can repeat the process
along all the other directions in the Cartan sub-algebra until the resulting
twelve fermion term, let us call it ${\widetilde f}^{(12)}$, is pluri-harmonic. 
The second step involves operators $Q_{a1}^*$, where 
\eqn\newop{ Q_{aA}^* = 
\g^j_{as}{\partial \over \partial \f_{sA}}{\partial \over \partial x^j_A},}
with no sum on $A$. These operators anti-commute with each other and reduce
fermion number.
On pluri-harmonic forms, these operators also anti-commute with 
$Q_b$. 

Now we apply the operators $Q_{a1}^*$ and $Q_{a2}^*$ to 
${\widetilde f}^{(12)}$ until we obtain a new $K$ fermion term $h^{(K)}$ which 
is 
killed by all of the operators $Q_a,Q_{aA}^*$, with respect to 
any choice of coordinates for the Cartan.  
This term is in general no longer invariant 
under $Spin(9)$. 
By our earlier discussion, we see that $K \geq 8.$  
We expand $h^{(K)}$ as before: 
$$h^{(K)} = \sum_i h^{(K)}_i.$$
We shall first show that $i\leq 8$. Note that $h^{(K)}_{i_{max}}$ is killed by,
$$\g^j_{as}\f_{s1}{\partial \over \partial x^j_1},$$ 
and by, 
$$\g^j_{as}{\partial \over \partial \f_{s1}}{\partial \over \partial x^j_1}.$$
If we let $r_i = |x_i|$ then,
\eqn\longcomp{ \eqalign{ 0 &= (x^k_1\g^k_{at}{\partial 
\over \partial \f_{t1}}\g^j_{as}\f_{s1}{\partial 
\over \partial x^j_1} - 
x^k_1\g^k_{at}\f_{t1}\g^j_{as}{\partial \over \partial \f_{s1}}{\partial \over 
\partial x^j_1})h^{(K)}_{i_{max}} \cr
&= \big[ ({\partial \over \partial \f_{s1}}\f_{s1}
 - \f_{s1}{\partial \over \partial \f_{s1}})r_1{\partial \over \partial r_1} +
  \g^{kj}_{ts}({\partial \over \partial \f_{t1}}\f_{s1} - 
{\partial \over \partial \f_{t1}}\f_{s1})x^k_1{\partial \over \partial 
x^j_1} \big] h^{(K)}_{i_{max}} \cr 
& = (16-2i_{max})r_1{\partial \over \partial r_1}
h^{(K)}_{i_{max}}.}} 
So 
$h^{(K)}_{i_{max}}$ is constant in $r_1$ and therefore zero unless $i_{max} = 
8$. 
We can apparently increase $i_{max}$ 
by acting on  $h^{(K)}_{8}$ with 
$$L= \f_{s1}{\partial \over \partial \f_{s2}}.$$ This operation 
corresponds to an infinitesimal coordinate change in the Cartan.
Since we have shown that
$i_{max}=8$ cannot in fact be increased, we can conclude that $L h^{(K)}_8 = 0.$ 
Anti-commuting $L$ and $Q_{a1}^*$ gives 
$\g^j_{as}{\partial \over \partial \f_{s2}}{\partial \over \partial x^j_1},$
which must also kill $h^{(K)}_8$.  
It is easy to see that $h^{(K)}_8$ is then also killed by,
$$x^k_1\g^k_{at}\f_{t2}\g^j_{as}{\partial \over \partial \f_{s2}}{\partial 
\over \partial x^j_1} =  
(K-8)r_1{\partial \over \partial r_1}  + 
\g^{kj}_{ts}\f_{t2}{\partial \over \partial \f_{s2}}x^k_1{\partial \over 
\partial x^j_1}.$$ 
In a similar way, we can deduce from the relation $Q_{a2}^*h^{(K)}_{i_{max}} = 
0$
that:
$$ \big[ (K-8)r_2{\partial \over \partial r_2}  + 
\g^{kj}_{ts}\f_{t2}{\partial \over \partial \f_{s2}}x^k_2{\partial \over 
\partial x^j_2} \big] h^{(K)}_8 = 0.$$ 
Combining these relations gives, 
\eqn\prefinal{ \big[ (K-8)r{\partial \over \partial r}  + 
\g^{kj}_{ts}\f_{t2}{\partial \over \partial \f_{s2}}(x^k_1{\partial \over 
\partial x^j_1} + x^k_2{\partial \over \partial x^j_2}) \big] h^{(K)}_8 = 0.}
It is convenient to rewrite this relation in the following form,
\eqn\finalform{ \big[ (K-8)r{\partial \over \partial r}   
+ (r_1+r_2)(C) - r_1(C) + (s+r_2)(C) - s(C)\big] h^{(K)}_{8,s}=0.}
We define $s$ as follows: applying the $Q_{aE}^*$ operators sent 
the $Spin(9)$ invariant $f^{(12)}$ into $h^{(K)}$ which is generally not 
$Spin(9)$
invariant. We can decompose $h^{(K)}$ into components which satisfy 
an equation of the form,
$$x^i_E {\partial \over \partial x^j_E} - x^j_E {\partial \over \partial x^i_E} 
+ 
\g^{ij}_{ts}\f_{tE}{\partial \over \partial \f_{sE}} + s(v^{ij}) h^{(K)}_{i,s} = 
0.$$
Here $s$ is a representation of $Spin(9)$ and the $v^{ij}$ are  generators 
of $Spin(9)$.  For example, if $K=11$ then $s$ is the spinor representation.
When $K = 12 - c$, $s$ is an irreducible representation appearing in the 
$c^{th}$ power of the spinor representation. 
 
We are left again with a Casimir term,  
$(r_1+r_2)(C) - r_1(C) + (s+r_2)(C) - s(C)$.
From representation theory, we learn that on an irreducible 
representation with highest $r_j$ weight of $l_j$ and highest $s$ weight of 
$l_3$, this term is bounded above by: 
\eqn\bounded{ \eqalign{& (l_1+l_2+2\delta,l_1+l_2) - (l_1+2\delta,l_1) + 
(l_2+l_3+2\delta,l_2+l_3)  \cr
& - (l_3+2\delta,l_3) = 2(l_1,l_2) + 2(l_2+2\delta,l_2) + 2(l_3,l_2).}}
Our prior computations imply that $l_1 = 4w_1$. Now we can work case by case:  
when $K=12$, we have $l_3 = 0$ and $l_2 \leq 2w_1+2w_2.$  
This gives a Casimir term no bigger than $80$ which implies 
homogeneity $\geq - 20$. This corresponds to a two-loop correction. 

When $K=11,$  $2l_2\leq 3w_1+3w_2+w_3+w_4$ 
and $2l_3\leq w_1+w_2+w_3 + w_4.$ Therefore the Casimir term is bounded by, 
$$ \eqalign{ & (4w_1,3w_1+3w_2+w_3+w_4) + {1\over 2}
(17w_1+13w_2+7w_3+3w_4,3w_1+3w_2+w_3+w_4) \cr
& + {1\over 2}(w_1+w_2+w_3+w_4,3w_1+3w_2+w_3+w_4) = 66.}$$
This implies homogeneity $\geq -22$ for $h^{(11)}$ and therefore 
homogeneity $\geq -21$ for the corresponding part of the
twelve fermion term $f^{(12)}$. 
If we assume, as is physically reasonable, that the effective action is 
analytic in the coupling constant then this term is again generated at two-loops
at worst. 

When $K=10$,  $l_2\leq w_1+w_2+w_3$ 
and $l_3\leq w_1+w_2+w_3+w_4.$  The Casimir term is now bounded by 
$$ \eqalign{ & (4w_1,2w_1+2w_2+2w_3) + (8w_1+6w_2+4w_3+w_4,2w_1+2w_2+2w_3)
 \cr & + 2|w_1+w_2+w_3|^2 = 50.}$$ 
This gives homogeneity $\geq -25$ for $h^{(10)}$ 
and $\geq -23$ for the corresponding part of $f^{(12)}.$ This is a potential
three-loop term. 

When $K=9$, 
the Casimir term is bounded by $25.$ This implies 
homogeneity $\geq -22$ for the corresponding term in $f^{(12)}$. Again, this is
at worst two-loop assuming analyticity in the coupling constant.  

When $K=8$, we have the additional equations, 
$$\g^j_{as}\f_{s1}{\partial \over \partial x^j_2} h^{(K)} = 0.$$
In a familiar way, we can extract the following relation:
$$ (8r{\partial \over \partial r}  - 
r_1(v^{kj})(r_1+s)(v^{kj})) h^{(K)}_{8,s} = 0.$$
In this case, the Casimir term is bounded by 
$$ \eqalign{ (r_1+s)(C) + r_1(C) - s(C) & \leq 2r_1(C) + 2(r_1,s) \cr
& =104. }$$
This gives homogeneity $-9$ for the corresponding term in
$f^{(12)}$, which must then come with an unphysical negative power of 
the coupling constant.

The one problematic case is $K=10$ which could be generated at three
loops. We will now show that this potential three-loop term cannot 
occur.  In obtaining the three-loop bound, we assumed that the original
twelve fermion term was pluri-harmonic. If it were not pluri-harmonic,
we would have to apply at least one Laplacian to get a pluri-harmonic
term. If this were the case, then our homogeneity bound should be raised
by $2$ which gives at most a two-loop correction, assuming analyticity
in the coupling.

 So let us consider a pluri-harmonic, Weyl invariant twelve 
fermion term  $f^{(12)}_i$. Since we are restricting to the $N=3$ case,
we have two scalar fields in the Cartan which we will label $x^j$ and $y^j$.
We can expand  $f^{(12)}_i$ in spherical 
harmonics as a sum of terms of the form: 
$$ x^I y^J |x|^{-7-2|I|} |y|^{-7-2|J|}.$$ 
The multi-indices $I$ and $J$ are to be contracted with indices on an 
appropriate
twelve fermion structure. 
The constraint equations then imply the existence of a term in $f^{(12)}_{12-i}$ 
of the form  
$x^L y^M|x|^{-7-2|L|} |y|^{-7-2|M|},$ where $|L| = |I|+12-2i$ and
$|M| = |J|-12 + 2i.$ This implies that $|I| \geq 2i-12.$ 

A three-loop term has homogeneity $-23$ and so $14 +|I|+|J| = 23,$ 
or $|I|+|J| = 9$.  This equality is incompatible with the previous 
bound on $|I|$ when $i>10$.  So, we need only consider 
$i_{max}=10,9$ and $8$. 

We will eliminate these possible terms in a way which can also be used to 
get {\it lower bounds} on loop corrections. 
Recall that $f^{(12)}_{i_{max}}$ is killed by the operators 
$\g^j_{as}\f_{s1}{\partial \over \partial x^j_1}.$ We then have the weaker 
equation, 
\eqn\newweak{ \left[ (16-i_{max})r_1 {\partial \over \partial r_1} + 
 \g^{kj}_{ts}{\partial \over \partial \f_{t1}}\f_{s1}x^k_1{\partial 
\over \partial x^j_1} \right] f^{(12)}_{i_{max}} = 0.}
We have not used this equation for $N>2$ because our invariance 
condition does not allow us to compute the Casimir term in general.
However, we can estimate it. 
Using $Spin(9)$ invariance, we rewrite this equation
in the form: 
$$ \left[ (16-i_{max})r_1 {\partial \over \partial r_1} + 2\rho_1(C) -
 \g^{kj}_{ts}{\partial \over \partial \f_{t1}}\f_{s1}x^k_2{\partial 
\over \partial x^j_2}\right] f^{(12)}_{i_{max}} = 0.$$ 
On terms in $f^{(12)}_i$ of the form 
$x^Iy^J|x|^{-7-2|I|}|y|^{-7-2|J|},$ the operator
$\g^{kj}_{ts}{\partial \over \partial \f_{t1}}\f_{s1}x^k_2{\partial 
\over \partial x^j_2}$, is bounded above by 
$|J|(16-i).$
The first Casimir term, $2\rho_1(C)$ is bounded above by 
$88$ for $i_{max} = 8,$ 
$84$ for $i_{max} = 9$ and
$60$ for $i_{max} = 10.$ 
We therefore see that the $x_1$ homogeneity given by $ -7-|I|$ is bounded below 
by 
$-11-|J|$ for $i_{max} = 8,$
$-12-|J|$ for $i_{max} = 9$ and
$-10-|J|$ for $i_{max} = 10.$ 
Using $|I|+|J| = 9$, $|I|$ is bounded above by 
$13-|I|$ for $i_{max} = 8,$
$14-|I|$ for $i_{max} = 9 $ and
$12-|I|$ for $i_{max} = 10.$ 
This implies $|I| \leq 7$. From this bound, we can conclude that
 $i_{max} = 10$ leads at most to 
a two-loop correction. Moreover, our constraints imply that for 
$i_{max} = 9,$ $(|I|,|J|) = (7,2)$ or $(6,3).$ 
Weyl invariance then implies that there must also be inadmissible 
solutions of the form $(|I|,|J|) = (8,1)$ or $(9,0)$. 
So $i_{max} = 9$ leads to at most a two-loop correction. 
A similar argument shows that $i_{max} = 8$ also 
leads to at most a two-loop correction. Therefore for $N=3$, the twelve fermion
terms are two-loop exact.

\newsec{Generalized F-terms}

Our technique for finding supersymmetry constraints can be summarized as
follows: consider the terms at a given order in the velocity expansion 
with the largest number of fermions. Let us denote these `top forms'
collectively by $ f^{(p)}$. In the simplest case, $ f^{(p)}$ only contains
fermions and no spacetime derivatives. Using the lowest
order free-particle supersymmetry transformations, $ f^{(p)}$ varys into a 
piece with one additional fermion and a piece with one fewer fermion. In 
searching for constraints, we generally want to restrict our attention to
the piece with one additional fermion. 

With this restriction,  we can think
of the supercharges $Q_a$ as differential operators acting as, 
\eqn\restrict{ Q_a = \g^i_{ab} \f_{bA} p^i_A. }
The restriction \restrict\  neglects the supersymmetry variation of the 
fermions.
The supersymmetry constraints then follow from the sixteen equations, 
\eqn\reviewcon{ Q_a f^{(p)} = S_a.}
The $S_a$ are source terms obtained by varying lower order terms in the 
Lagrangian
using corrections to the supersymmetry transformations, encoded in $N,M$
of \newtransforms. Typically, only variations of lower order top forms appear
in $S_a$. 

Our theory always contains at least sixteen fermions so we always have top forms
in the supersymmetric completion of the $v^4$ and $v^6$ terms. As we have seen, 
the equations \reviewcon\ are strong enough to completely determine the coupling
constant dependence of the eight fermion terms for any $N$. For $N=3$, the same
is true for the twelve fermion terms but this is no longer clear for higher $N$.
For higher $N$, we should expect that only certain terms at a given order in the
velocity expansion will be constrained by \reviewcon. We gave an example of 
such a term in section $3.1$. 

Let us return momentarily to the $SU(2)$ case as an example. The top form in 
the supersymmetric
completion of $v^8$ is a sixteen fermion term. There is a unique structure that
takes the form,
\eqn\sixteen{ f^{(16)} = f(r) \, \f_{a_1} \cdots \f_{a_{16}} 
\epsilon^{a_1\cdots a_{16}}, }
for some radial function $f$.  It is easy to check that $ f^{(16)}$ can be 
written
in the form, 
\eqn\Dterm{ f^{(16)} = \left\{ Q_1, \cdots  \acom{Q_{16}}{ g(r)} \right\}, }
for some $g(r)$, obtained by appropriately integrating $f(r)$. The brackets
appearing in \Dterm\ should be viewed as graded commutators  i.e. 
anti-commutators
for two fermionic operators and commutators for everything else. Now the 
variation
of \sixteen\ into a term with seventeen fermions automatically vanishes since
we only have sixteen fermions. From \reviewcon, we find no constraint on the 
choice of $g(r)$.  

Top forms that can be written in the form \Dterm\ are natural generalizations of 
the superspace notion of a D-term. While there is no useful notion of superspace
for theories with sixteen supercharges, it is still meaningful to ask whether
 a term in the Lagrangian -- not necessarily a top form -- can be written in the 
form
\Dterm. Top forms that are generalized D-terms in this sense are automatically
killed by each $Q_a$. In searching for supersymmetry constraints, we want to 
quotient out by these trivial solutions in the usual cohomological sense.  
It is then natural to define the set of 
generalized F-terms as all terms that cannot be written in the form 
\Dterm.\foot{We
wish to thank E. Witten for suggesting this definition.} 
Since for $SU(2)$ all top forms at order eight in the
derivative expansion are clearly D-terms, we do not expect supersymmetry to 
impose 
any simple restrictions on terms of this order. 

However for higher rank, we have more fermions so there can be generalized 
F-terms 
at higher orders in the derivative expansion of the most general
effective action compatible with sixteen supersymmetries and the global 
$Spin(9)$
symmetry. A strong indication of the existence of such terms is the agreement of
certain interactions in Matrix theory with supergravity found in \rdinetwo. We
would hope that there is a simple argument showing that the agreement is because
of non-renormalization theorems. This is an important open question. 

We will conclude
our discussion by showing that there are possible 
generalized F-terms in the supersymmetric
completion of $v^8$ terms for $N>3$. The argument goes as follows: we can choose 
a
basis for the gamma matrices $\g^i$ so that $\g^9$ is diagonal while the rest 
are
of the form, 
$$  \pmatrix{ 0 & A \cr A^T & 0 \cr}, $$
where $A$ is some $8\times 8$ matrix. So $p^i \g^i_{ab} \f_{b}$ can contain
the fermions,
$$ \f_1, \ldots, \f_8, \f_a$$
for $a\geq 9$. We are suppressing the Cartan labels for the moment and just 
focusing
on the $Spin(9)$ indices. For $a<9$, we see that $p^i \g^i_{ab} \f_{b}$ can 
contain
the fermions,
$$ \f_a, \f_9, \ldots, \f_{16}. $$
So, for example, $\f_1$ can only occur for nine choices of $a$. 
We could then consider a top form, 
$$  \f_{1 A_1} \cdots \f_{1 A_{16}}, $$
which is possible if the rank of the group is sixteen or larger. This term, 
which can 
appear as part of some $Spin(9)$ invariant top form, clearly cannot
be written as a generalized D-term of the form \Dterm. We can find analogous 
examples 
in theories with lower rank. Let us take rank three:
a sixteen fermion term with the following fermion content, 
$$  \left( \f_{1 A_1} \f_{1 A_2}\f_{1 A_3}\f_{2 A_1}\f_{2 A_2}\f_{2 A_3}
\cdots \f_{5 A_1}\f_{5 A_2}\f_{5 A_3} \right) 
\times \f_{6 A_3}, $$
cannot be part of a generalized D-term. Only fourteen $Q_a$ could possibly 
contribute
$ \f_{1}, \ldots, \f_{6}$ fermions but we need sixteen such fermions. It is not 
hard
to construct more examples of this kind. 

\bigbreak\bigskip\bigskip\centerline{{\bf Acknowledgements}}\nobreak
It is our pleasure to thank M. Dine, Y. Kiem, D. Lowe, S. Paban and 
E. Witten for helpful conversations.  The work of 
 S.S. is supported  by the William Keck Foundation and by 
NSF grant PHY--9513835; that of M.S. by NSF grant DMS--9870161.

\vfill\eject

\listrefs
\bye